# Cryogenic Calibration Setup for Broadband Complex Impedance Measurements

P. Diener[a,b], F. Couëdo[a], C. Marrache-Kikuchi[a], M. Aprili[b] and J. Gabelli[b]

[a] *Centre de Sciences Nucléaires et de Sciences de la Matière, Université Paris Sud, 91405 Orsay Cedex*
[b] *Laboratoire de Physique des Solides, Université Paris Sud, 91405 Orsay Cedex*

**Abstract.** Reflection measurements give access to the complex impedance of a material on a wide frequency range. This is of interest to study the dynamical properties of various materials, for instance disordered superconductors. However reflection measurements made at cryogenic temperature suffer from the difficulty to reliably subtract the circuit contribution. Here we report on the design and first tests of a setup able to precisely calibrate in situ the sample reflection, at 4.2 K and up to 2 GHz, by switching and measuring, during the same cool down, the sample and three calibration standards.

## INTRODUCTION

To study quantum phase transitions (QPT), it is necessary to measure the absolute value of the complex impedance at finite frequency and very low temperature. Indeed, in the quantum regime $k_B T \ll \hbar\omega$, the transport properties of a system close to a QPT should exhibit critical fluctuations which are governed by $\hbar\omega$ [1, 2].

One example of QPT is the superconductor-to-insulator transition observed in thin films when increasing the disorder. Probing the complex impedance in disordered thin films can thus provide clues to the relative evolution of the superfluid density, mainly governed by the imaginary part of the impedance, and the quasiparticle response, mainly accessible through the real part of the impedance [3].

One can measure the complex impedance $Z(\omega)$ on a wide frequency range for instance by probing the reflection $S_{11}(\omega)$ of an alternating GHz signal applied to the sample via a transmission line. To fully calibrate the setup, it is necessary to measure the reflection of three samples with known impedances in addition to the sample under study, with the circuit being under the same conditions of temperature, power and field. This procedure, simple at ambient temperature, becomes challenging at cryogenic temperatures, because one should iteratively cool and warm the cryostat to change the sample, thus potentially changing the transmission line response, whereas the measurements should, in principle, be made in the exact same conditions [4-6].

We have developed a unique calibration setup designed for microwave frequencies which allows switching in situ between the four samples at liquid helium temperature. These results validate the calibration setup performance in a cryogenic environment.

## PROCESS OF CALIBRATION IN MICROWAVE REFLECTOMETRY

The calibration procedure used here is the standard one port error model commonly used to calibrate commercial vector network analyzers. The setup errors are described by three error terms, as shown Figure 1: the directivity $E_D$, the reflection tracking $E_R$ and the source mismatching $E_S$. They are usually frequency dependent complex numbers.

The measured reflection of the sample placed at the end of the circuit transmission line $S_{11}^{meas}(\omega)$ is related to the actual sample reflection $S_{11}(\omega)$ by the relation:

$$S_{11}^{meas}(\omega) = E_D(\omega) + \frac{E_R(\omega) S_{11}(\omega)}{1 - E_S(\omega) S_{11}(\omega)}$$

and the actual sample reflection is related to the complex impedance of the sample by $S_{11} = (Z - Z_0) / (Z + Z_0)$, with $Z_0$ the characteristic impedance of the line, which is typically 50 Ω.

For an ideal system, $E_D = E_S = 0$ and $E_R = 1$ for all frequencies. This is not true for a real system having attenuation, parasitic resonances and delays. In addition, for cryogenic measurements, a part of the circuit is necessarily placed inside the cryostat, with the sample, giving rise to significant changes of the circuit properties and the error terms.

To determine $E_D$, $E_R$ and $E_S$, three samples of known impedances $Z_1$, $Z_2$ and $Z_3$, the calibration standards, are measured with the same circuit. With $a_i = (Z_i - Z_0) / (Z_i + Z_0)$ the actual reflection of the $i^{th}$ calibration standard and $m_i$ the corresponding measured reflection, we obtain a set of three (complex) equations to determine the three (complex) error terms, thus giving [5]:

$$E_D(\omega) = \frac{m_1(m_2 - m_3)a_2a_3 + m_2(m_3 - m_1)a_3a_1 + m_3(m_1 - m_2)a_1a_2}{(m_1 - m_2)a_1a_2 + (m_2 - m_3)a_2a_3 + (m_3 - m_1)a_3a_1}$$

$$E_R(\omega) = \frac{(m_1 - m_2)(m_2 - m_3)(m_3 - m_1)(a_1 - a_2)(a_2 - a_3)(a_3 - a_1)}{[(m_1 - m_2)a_1a_2 + (m_2 - m_3)a_2a_3 + (m_3 - m_1)a_3a_1]^2}$$

$$E_S(\omega) = \frac{m_1(a_2 - a_3) + m_2(a_3 - a_1) + m_3(a_1 - a_2)}{(m_1 - m_2)a_1a_2 + (m_2 - m_3)a_2a_3 + (m_3 - m_1)a_3a_1}$$

In the calibration process, it is important to define properly the reference plane: the calibration standards are measured at the exact equivalent circuit termination, with the same connectors and cable length. Indeed, a phase difference or an additional loss would give rise to a change of the circuit error terms between the three measurements, thus potentially breaking the calibration model. This is why we have chosen a circular symmetry for the four terminations, as described in the next section.

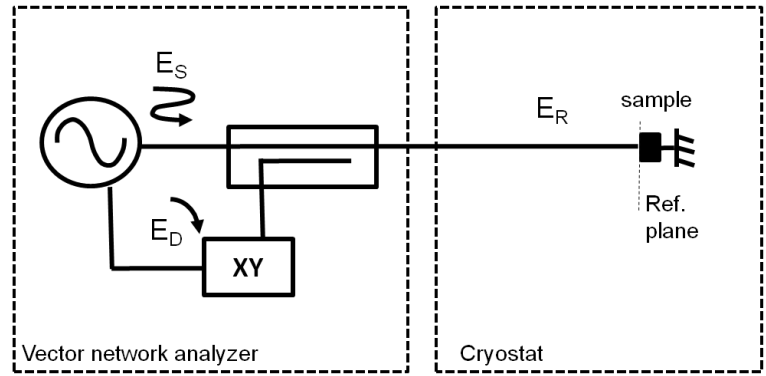

**FIGURE 1.** Physical signification of the calibration error terms in a standard one port reflectometry measurement. The directivity comes from the limited insulation of the directional coupler inside the vector network analyzer, the source mismatching relates to the parasitic signal coming back from the generator after sample reflection and the reflection tracking is due to the insertion loss of all the circuit cables and connectors. The dotted line indicates the reference plane of the calibration.

## CRYOGENIC CALIBRATION SETUP

The setup is presented Figure 2 and 3. The three calibration standards and the sample are placed at the end of four identical planar transmission lines, each of which is 6.6mm long, placed at 90° of each other, in a dial configuration. The signal enters in the reflectometer through a rigid copper coaxial cable oriented perpendicular to the dial and terminated by a right angle SMP/SMT Radiall connector, the selector. By rotating the copper cylinder containing the input coaxial cable, the selector, having a 380 μm core width, can be aligned to one of the four lines of width 470 μm, thus selecting the sample under test.

The rotation of the selector, the input coaxial cable and the cylinder is activated with the drive shaft placed in parallel and connected by gears. The MMPX Huber Suhner connectors having a slide-on plug are used to allow the rotation of the coaxial cable connected to the needle while keeping the rest of the RF circuit fixed. Special care is given to the upward coaxial cable fixation in order to have a reproducible connector to connector contact after each rotation.

The alignment of the selector to the on chip transmission lines is obtained using a mechanism of four ball spring plungers and four linear notches along the central rotating cylinder: as shown in Figure 2, when the rotation is activated, the cylinder has four more favorable positions, corresponding to the balls aligned to the notches.

The dial is made of a Rodgers 635 μm thick TMM10i chip covered by 17 μm thick copper on both sides. The circuit is drawn using a CNC milling machine.

To insure a robust selector/dial contact, special care is given to have a good flatness of the two pieces, by filling the gaps on both surfaces with epoxy and polishing them carefully.

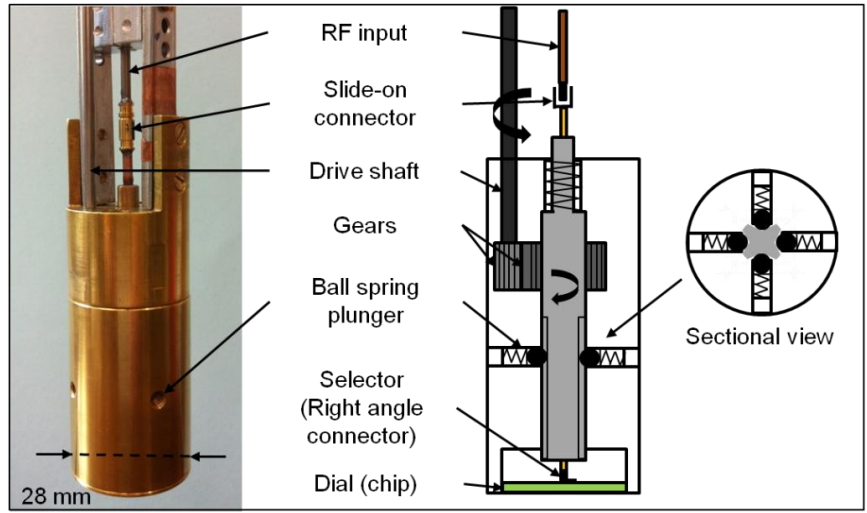

**FIGURE** 2. Picture and cross-section view of the calibration setup.

The details of the dial design are presented Figure 3. The microwave lines are conductor backed coplanar wave guides (CBCPW) of l = 6.6 mm length, s = 470 μm width and w = 220 μm gaps. The top and bottom groundplanes are connected by via holes filled with silver paste. Additionally, the top groundplane is connected to the box side by millimeter spaced aluminum bounding wires. As pictured in Figure 3, the calibration standards used are: 1) a line terminated by a gap g = 220 μm 2) a short circuited line and 3) a $R_{50}$ = 50 Ω Vishay Sfernice resistor of 480 μm long and 390 μm large, soldered on top of a 220 μm gap identical to the gap of the open circuit. The sample used to test the setup is another resistor having R=200 Ω. The reference plane is indicated in the figure. For simplicity, the standards will be called open, short and 50 Ω standards in the following. As shown in the equivalent circuits Figure 3, there are additional nonzero capacitance $C_{OC}$ for the open circuit gap and inductance $L_{SC}$ for the short circuited line to be taken into account. Thus, the actual impedances are $Z_1 = (i\omega C_{OC})^{-1}$, $Z_2 = i\omega L_{SC}$ and $Z_3 = R_{50}Z_1/(R_{50}+Z_1)$. The parasitic capacitance and inductance are estimated by [7]:

$$C_{OC} = \frac{(s+2w)}{4} \frac{\sqrt{\varepsilon_{eff}}}{z_0 c}$$

$$L_{SC} = \frac{2}{\pi} \varepsilon_0 \varepsilon_{eff} (s+w) z_0^2 [1 - \frac{1}{\cosh(60\pi^2/z_0\sqrt{\varepsilon_{eff}})}]$$

with c the speed of light in vacuum, $\varepsilon_0$ the vacuum permittivity and $\varepsilon_{eff}$ the effective dielectric constant. $\varepsilon_{eff}$ = 5.37 is determined with the microwave designer software AppCAD and we obtain $C_{OC}$ = 40 fF and $L_{SC}$ = 40 pH in our case.

The setup is mounted on the cold plate of a helium 4 bath cryostat. Most of the mechanical parts of the setup are in copper, and a vertical spring is placed on top of the central cylinder to maintain the selector in good contact with the dial despite the thermal contractions. The drive shaft is extended up to a wheel placed outside the cryostat which is activated manually.

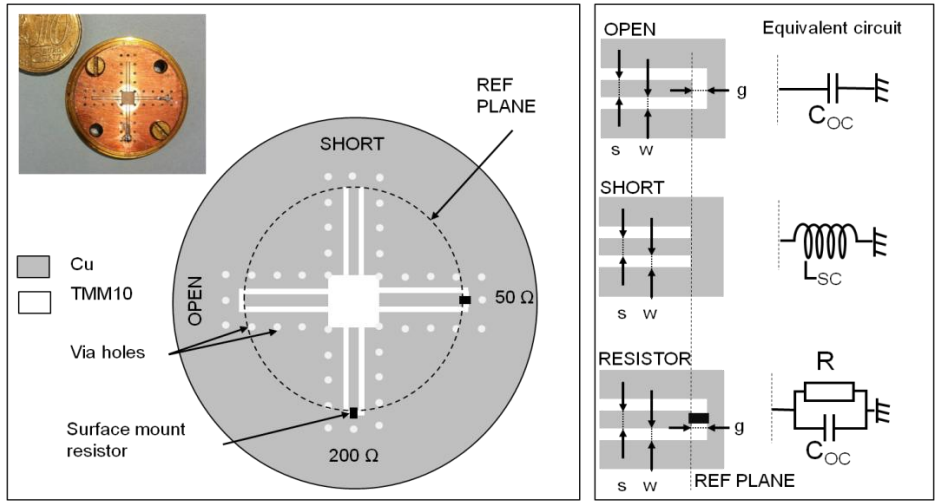

**FIGURE 3.** Left: Picture and top view of the dial. Right: calibration standard design and equivalent circuit. The reference plane is indicated by a dotted line.

## RESULTS

The calibration setup has been tested at ambient temperature and 4.2 K. Figure 4a shows the typical reflection magnitude obtained between 10 MHz and 2 GHz, with a microwave input power of 1 mW, when the selector is successively aligned to each three standards and to the test sample. Similar results have been obtained at the two temperatures. At the lowest frequency, the measured reflections are close to the ideal case where:

- $S_{11}^{meas}$ (open) = 1 ( total reflection, $|S_{11}^{meas}$ (open)$|$ = 1 and $\angle S_{11}^{meas}$ (open) = 0 )
- $S_{11}^{meas}$ (short) = -1 ( total reflection, $|S_{11}^{meas}$ (short)$|$ = 1 and $\angle S_{11}^{meas}$ (short) = $\pi$)
- $S_{11}^{meas}$ (50 $\Omega$) = 0 (total transmission).

The exponential magnitude decay with frequency (linear decay in db scale), clearly seen in Figure 4b, is related to the skin depth in the coaxial cables and is lower at low temperature due to the conductivity increase. In phase, the delay difference between the two temperatures, corresponding to the different slopes, is related to the thermal contraction of the coaxial cables.

The reproducibility is estimated by comparing successive measurements for each standard. The difference in magnitude between two measurements is less than 0.02 dB at ambient temperature and less than 0.1 dB at 4.2 K, indicating good mechanism stability.

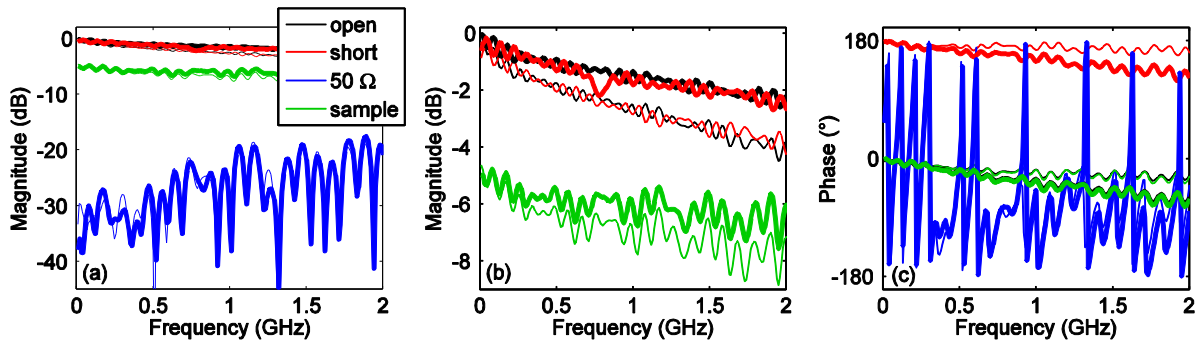

**FIGURE 4.** Magnitude (a) and phase (b) of the measured reflection of the three standards and the sample, at ambient temperature (thin lines) and 4.2 K (bold lines). (b) Zoom of (a) above -9dB to distinguish the signal increase with the cooling.

The measurements of the open, short and 50 Ω terminated lines are used to calibrate the test sample, following the procedure described above. The results obtained at 4.2 K are presented Figure 5.

As seen Figure 5a and b, the calibration correctly eliminates the attenuation, the ripples and the delay observed in the raw data and the corrected data show the typical flat reflection of a resistor.

The impedance is deduced from the corrected reflection. As in the case of the 50 Ω standard, there is a small parasitic capacitance $C_{OC}$ = 40 fF in addition to the desired sample resistance $R_{sample}$, which we eliminate from the impedance by inverting the relation $Z_{sample} = R_{sample}Z_1/(R_{sample}+Z_1)$.

The real and imaginary parts of $R_{sample}$ deduced are shown Figure 5c and d. It is compared to the measured DC value of the sample resistance, which is 185 Ω at 4.2 K. Within 2 Ω ripples, the result exhibits the expected flat and purely resistive impedance. There is still a 6 Ω spurious peak around 1 GHz observed in both real and imaginary part. This peak, related to a small reproducibility decrease at the same frequency (not shown), is probably due to a slight selector misalignment.

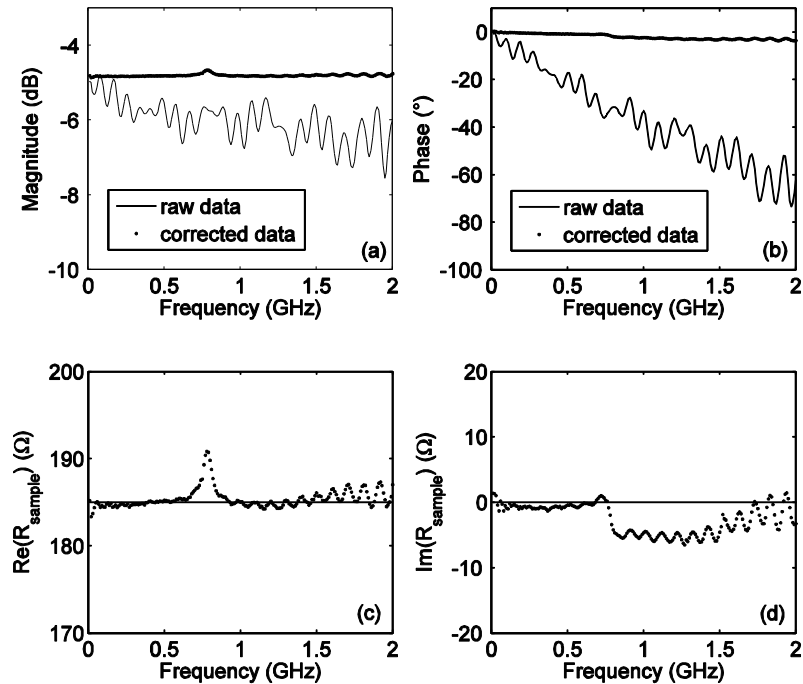

**FIGURE 5.** Top panel: magnitude (a) and phase (b) of the measured and calibrated sample reflection, at 4.2 K. Bottom panel: real (c) and imaginary (d) part of the sample resistance deduced from the corrected data (dot) and lines at $Re(R_{sample})$ = 185 Ω and $Im(R_{sample})$ = 0 for comparison.

Regarding the use of this setup for physical measurements near a QPT, the quantum limit is set by the ratio of the frequency to the temperature. It is therefore interesting to also check the calibration setup at higher frequency. We have tested the setup in the frequency range DC-20 GHz. The results are presented Figure 6. In addition to the skin depth decrease, we observe an increase of the 50 Ω reflection with frequency, due to the vector network analyzer limitation. Around 10 GHz, We have still $|S_{11}^{meas}(50\ \Omega)| \ll |S_{11}^{meas}(open)|, |S_{11}^{meas}(short)|$ which indicate that the circuit is still truly passing and could be used at least up to 10 GHz.

However, two shallow parasitic deeps are observed around 3 and 7 GHz. We attribute these deeps to the resonances inside the 24mm dial box. Dial box resonances cannot be subtracted by the calibration model, because they are not detected identically for each on chip transmission lines. This hypothesis is confirmed by looking at the calibration result, also Figure 5. The skin depth attenuation seen in the raw data is correctly subtracted, but two broad peak features appear in the corrected magnitude around 3 and 7 GHz which are incompatible with a simple resistor signature. To eliminate the dial box resonances, one can reduce the dial box size, thus pushing the resonant modes at higher frequency, and add radiation absorbing material on the inner walls of the box.

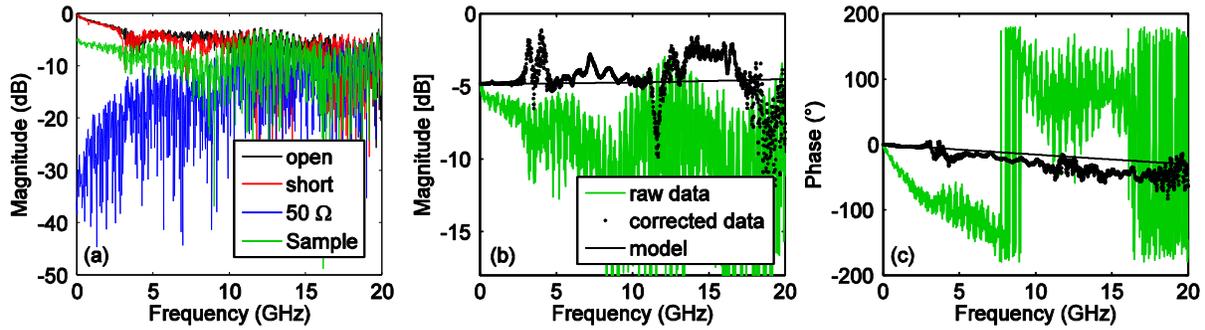

**FIGURE 6.** Typical calibration result on the frequency range DC-20 GHz. (a) $|S_{11}^{meas}|$ for the three calibration standards and the sample. (b) and (c) Magnitude and phase reflection of the sample before and after calibration, compared to the result expected by the RC model with $R_{sample} = 185\ \Omega$ and $C_{OC} = 40$ fF.

These results, at 4.2 K, validate the calibration setup performance also at lower temperature: the main performance limitation related to the cooling is the mechanic changes due to the thermal contractions, which are negligible below 4.2 K. The current setup is thus fully compatible with a dilution refrigerator, allowing the study of the transport properties in the quantum regime.

## CONCLUSION

We have developed a cryogenic calibration setup for microwave reflection measurements. The first tests, made at ambient and liquid helium temperature, have shown that the complex impedance can be determined within few $\Omega$ in the range DC - 2 GHz.

## ACKNOWLEDGMENTS

We greatly acknowledge G. Guillier for his contribution to the mechanical design of the setup. This work has been partially supported by the Agence Nationale de la Recherche (ANR) (Grant No. ANR-2010-BLANC-0403-01).